%% file: going-live.tex
\documentclass[sigconf]{acmart}
\usepackage{url}
\usepackage{graphicx,epsfig}
\usepackage{subfig}
\usepackage{graphics,dcolumn,bm,epic, eepic,fleqn,float}
\usepackage{amssymb,amsmath,multirow,rotate,color}
\usepackage{epstopdf}
\usepackage{soul}
\usepackage{listings}
\usepackage{color}
\usepackage{booktabs}
\usepackage[T1]{fontenc}
\usepackage{xspace}
\usepackage[flushleft]{threeparttable}
\usepackage{booktabs}
\usepackage{balance}
\usepackage{siunitx}

\newcommand{\head}[1]{\multicolumn{1}{c|}{#1}}

\renewcommand{\footnotesize}{\scriptsize}

\newcommand{\one}{({\em i}\/)}
\newcommand{\two}{({\em ii}\/)}
\newcommand{\three}{({\em iii}\/)}

\def\eg{\emph{e.g.,}\xspace}
\def\etc{\emph{etc.}\xspace}
\def\ie{\emph{i.e.,}\xspace}
\def\etal{\emph{et al.}\xspace}

\def\cf{\emph{cf.}\xspace}

\ccsdesc[500]{Social and professional topics~User characteristics}
\ccsdesc[300]{Networks~Network components}
\ccsdesc[300]{Networks~Network measurement}
\ccsdesc[500]{Information systems~World Wide Web}
\ccsdesc[300]{Information systems~Information systems applications}

\copyrightyear{2018}
\acmYear{2018}
\setcopyright{iw3c2w3}
\acmConference[WWW 2018]{The 2018 Web Conference}{April 23--27, 2018}{Lyon, France}
\acmPrice{}
\acmBooktitle{WWW 2018: The 2018 Web Conference, April 23--27, 2018, Lyon, France}
\acmDOI{10.1145/3178876.3186061}
\acmISBN{978-1-4503-5639-8/18/04}
\keywords{social live broadcast; user behaviour; edge computing}

\begin{document}
\title{Facebook (A)Live? Are live social broadcasts really \emph{broad}casts?}

\author{Aravindh Raman}
\affiliation{%
	\institution{King's College London}
}
\email{aravindh.raman@kcl.ac.uk}

\author{Gareth Tyson}
\affiliation{%
	\institution{Queen Mary University of London}
}
\email{g.tyson@qmul.ac.uk}

\author{Nishanth Sastry}
\affiliation{%
	\institution{King's College London}
}
\email{nishanth.sastry@kcl.ac.uk}

\renewcommand{\shortauthors}{Aravindh et al.}

\begin{abstract}
	
	The era of live-broadcast is back but with two major changes. First, unlike traditional TV broadcasts, content is now streamed over the Internet enabling it to reach a~\emph{wider} audience. Second, due to various user-generated content platforms it has become possible for~\emph{anyone} to get involved, streaming their own content to the world. This emerging trend of \emph{going live} usually happens via social platforms, where users perform live social broadcasts predominantly from their mobile devices, allowing their friends (and the general public) to engage with the stream in real-time. With the growing popularity of such platforms, the burden on the current Internet infrastructure is therefore expected to multiply. With this in mind, we explore one such prominent platform --- Facebook Live.  
	We gather 3TB of data, representing one month of global activity and explore the characteristics of live social broadcast. From this, we derive simple yet effective principles which can decrease the network burden. We then dissect global and hyper-local properties of the video while on-air, by capturing the geography of the broadcasters or the users who produce the video and the viewers or the users who interact with it. Finally, we study the social engagement while the video is live and distinguish the key aspects when the same video goes on-demand. 
	A common theme throughout the paper is that, despite its name, many attributes of Facebook Live deviate from both the concepts of \emph{live} and \emph{broadcast}.
\end{abstract}
\maketitle
\input{sections/introduction}
\input{sections/background}

\input{sections/datasets}

\input{sections/char}
\input{sections/geo}
\input{sections/social}
\input{sections/conclusions}

{
\balance{
	\bibliographystyle{ACM-Reference-Format}
	\bibliography{bib/biblio} 
	}
}

\end{document}

%% file: sections/introduction.tex
\section{Introduction}

The online sharing of user-generated content is now a day-to-day activity for many users around the world~\cite{cisco2016report}. Whereas traditionally this has followed a ``static'' upload model (where users create a video offline, and then upload their content on a platform like YouTube for on-demand viewing), recent years have seen a noticeable shift towards live equivalents. Mobile applications such as Facebook Live, Periscope and Instagram Live allow users to easily broadcast their activities to a potentially global audience. We term this medium \emph{social live broadcast}, and distinguish it from more traditional forms of live streaming such as Twitch, which are not based on social platforms and are not impacted by the various cultural and demographic implications of a social graph-based delivery.

One prominent example of live social broadcast is \emph{Facebook Live}. The platform is simple: it allows any Facebook user to ``go live'' and stream their mobile camera feed to friends (and others if public). An interesting twist is that the video optionally remains available after the broadcast, to be viewed on-demand, thus offering functionality similar to traditional user-generated platforms like YouTube. From a user behaviour perspective,  Facebook Live is a particularly powerful platform due to the presence of both amateur and professional users, and the combination of live and non-live video delivery. Further, with a strong social network and a global reach, Facebook allows us to extract social features and geographical patterns. From an infrastructural perspective, it also opens up several challenges and opportunities, \eg to exploit social localities for edge caching.

In this paper, we make two broad contributions. First, we characterise user behaviour in mobile live social broadcast. Second, where appropriate, we highlight relevant implications for mobile live social video delivery derived from our observations. To this end, we have collected a large-scale dataset covering one month of activity on Facebook Live. This dataset not only encompasses information regarding access patterns, but also geographical indicators on where broadcasters and viewers are located. To the best of our knowledge, we are the first to study Facebook Live and, importantly, the first to derive key systems insights. Whereas the limited body of past research has focussed on performance aspects~\cite{wang2016anatomy,siekkinen2016anatomy}, we strive to understand its end user characteristics, as well as how this might create architectural design opportunities.

We begin by characterising broadcaster and viewer properties (\S\ref{sec:char}). We quantify Facebook Live's popularity, observing 3.2 million accounts performing broadcasts, with a peak of 62 million online viewers (\S\ref{sec:char:pop}). We identify three types of broadcasts. The majority (95\%) are from social users on Facebook's mobile app, whilst the minority are from Facebook page accounts (4\%) and applications using Facebook's third party API (1\%). 
Although social users typically perform short broadcasts, we find a non-negligible number of mobile users with broadcast sessions exceeding one hour (\S\ref{sec:char:duration}). However, nearly half of these streams go unwatched, even for regular broadcasters. Despite this, we find that the Facebook app uploads all content regardless of viewers, wasting user device battery and network resources. We therefore propose a simple scheme, whereby content is locally cached on a device until viewers arrive.  We find that 21.91\% (21.33TB)  of data could be offloaded from delivery through this simple innovation (\S\ref{sec:char:necessary}).\looseness=-1

We then dissect the geographical patterns of both broadcasters and viewers (\S\ref{sec:geo}). We confirm that Facebook Live has a global reach (\S\ref{sec:geo:where}). The majority of popular broadcasters are based in Europe, South East Asia, Brazil and the East Coast of the United States. Driven by social properties, most broadcasts are highly localised. For example, 8\% of viewers are within a city-level radius of 25KM from the broadcaster, accounting for 1.57M viewers. 
That said, we also observe a significant proportion of international consumption too (overall, 39.8\% of viewers are domestic and the remaining are international). There are some distinct power players here. For example, residents in the United States stream a disproportionate amount of broadcasts to Mexico and Canada. This highlights well the social properties of Facebook Live, whereby users tend to consume directly from friends (who may be in different countries due to immigration).

Finally, we refocus our attention on user engagement (\S\ref{sec:social}). We find that popular broadcasts consistently garner views rapidly. However, this is not the norm --- the median view count typically remains below 10 throughout the entire duration of streams. We also compare this against archived on-demand videos, \ie streams that were live but subsequently made available for later viewing. We find that, in fact, most engagement (comments, likes, shares) happen \emph{after} a stream has been archived, thereby demoting the importance of the live component. We finally combine the above observations to question to what extent Facebook Live is truly live, social and broadcasting (\S\ref{sec:conclusions}).\looseness=-1

%% file: sections/background.tex
\section{Background \& Related Work}
 
\vspace{4pt}

\noindent\textbf{Online Video Delivery:} Our work overlaps with studies looking at user generated (UGC) video platforms, \eg~\cite{Ameigeiras2012,cha2009analyzing,tyson2013demystifying}. These include user behaviour analysis~\cite{Xing2011,joglekar2017socinfo} to understand consumption patterns~\cite{Yu2008,Hu2012,Gopalakrishnan2010,Chang2008,lee2017globecomm}. Facebook Live departs from the traditional UGC model of interaction, with users live broadcasting through their mobile device and online social network (rather than simply uploading videos). This allows for a number of new lines of exploration, particularly relating to how \emph{live} videos are created and consumed by social ties. 
There have been a small number of studies into live streaming platforms, including Akamai~\cite{sripanidkulchai2004analysis}, BBC iPlayer~\cite{karamshuk2015ssc2,karamshuk2016jsac,nencioni2016score,karamshuk2015ssc1} or CNLive~\cite{li2011measurement}. These professional platforms serve a wide range of international and professional content, although we emphasise that this differs dramatically from the user generated content discussed above. Perhaps the most famous user generated live streaming platform of the day is Twitch, which has seen increasing research attention~\cite{deng2017internet,hamilton2014streaming,deng2015behind}. 
This streams live computer gameplay, but differs from Facebook Live in that it is not mobile, nor integrated within a social network.

\vspace{4pt}

\noindent\textbf{Geographical Video Access Patterns:} Another major part of our work is investigating the geographical properties of social broadcast and consumption. There have been a small set of studies looking at the geographical properties of video access. Li \etal~ looked at the the location of viewers watching PPTV in China~\cite{li2014geographic}. They found that a significant portion of videos get at least 70\% of their views from just 3 out of 33 provincial locations. Similar studies have been performed for YouTube, finding that users tend to access videos nearby to them~\cite{brodersen2012youtube}. Whereas studies have been done looking at spatial localities in Facebook~\cite{wittie2010exploiting} and Twitter~\cite{rodrigues2011word}, this has not inspected broadcasts. Interestingly, the integration of live social broadcast with a platform like Facebook also conflates two factors impacting spatial access patterns: \one~Content locality, which is the propensity for content generated nearby to be more relevant to a user; and \two~Social locality, which is the propensity of content generated by friends to be more relevant.

\vspace{4pt}

\noindent\textbf{Social Live Broadcasts:} We take inspiration from the above works, but emphasise \emph{social live broadcasts platforms} such as  Facebook Live and Periscope~\cite{siekkinen2016anatomy,wang2016anatomy} are distinct from these systems in a number of ways. They allow any user to broadcast themselves (usually from a mobile device) to their friends and possibly other interested parties. Unlike traditional user-generated video platforms, live social streaming drives consumption in realtime. Critically, this is done through direct integration with a social network, notifying online users of the broadcast. 
This throws up interesting questions that do not emerge in past studies on live platforms. For example, we later explore the social interactions between viewers and broadcasters in real time. Beyond this, the unique social API of Facebook Live allows us to dive into questions regarding user locations to understand their impact on content creation and consumption. \looseness=-1

To the best of our knowledge, we are the first to study Facebook Live at scale; thus, intentionally, we strive to offer a broad characterisation covering the key novel aspects of the system. There are, however, a small number of studies on related social live streaming platforms. Wang \etal studied Periscope from an infrastructural perspective~\cite{wang2016anatomy}, focussing on understanding how Periscope scales up. This focussed on popular streams, whereas we also focus on the behaviour of unpopular streams (which are more frequently occurring). We also integrate this with exploration of geographical and social properties of the broadcasters. A related paper~\cite{siekkinen2016anatomy} inspected Periscope to capture performance metrics, \eg video quality, stall events, power consumption. Tang \etal focussed on a small set of Meerkat and Periscope broadcasts using crowd sourced analysis and interviews; they focussed on things like motivation for usage~\cite{tang2016meerkat}. This social analysis, however, was based on just 767 (manually analysed) live broadcasts. The focus of our paper is very different. We take a user-facing perspective, empirically exploring the ways users interact with Facebook Live on both a global and regional level. Although we do this at scale, we also focus on both highly popular and unpopular users to understand system implications for both.

%% file: sections/datasets.tex
\section{Datasets \& Methodology} \label{sec:dataset}
To explore Facebook Live, it is first necessary to collect a large and meaningful dataset of both the broadcasters and viewers. Hence, we begin by describing the data collection methodology employed.

\begin{table*}[t]
	\centering
	\caption{Summary of dataset separated across continents. Peak view counts are computed cumulatively across all videos in a snapshot (originating on the continent).}
	\begin{threeparttable}
		\begin{tabular}{| l | r | r | r | r | r | r | r|}
			\hline
			
			Attribute & \head{Total} & \head{NA} & \head{EU} & \head{AS} & \head{AF} & \head{SA} & \head{OC} \\
			\hline\hline
			\# of broadcasts  & 6.5 M & 1.9 M & 707 K  & 2.66 M & 117 K & 810 K & 52 K\\
			\# of broadcasters & 3.29 M  & 856 K & 336 K & 1.35 M & 64.6 K & 418 K &  25.4 K \\
			\hline
			\# of broadcasts with over 100 views &   61 K & 15.2 K & 8.2 K & 23 K & 1.5 K &  6.4 K & \head{426}  \\
			\# of broadcasters with over 100 views &  19 K & 4.8 K & 3 K & 6.5 K & \head{522} & 2.5 K & \head{163}  \\
			\hline
			Peak \# of viewers & 60 M & 17.6 M  &  7.39 M & 20 M & 1 M & 6.8 M & 387 K \\
			\hline
		\end{tabular}
	\end{threeparttable}
	\label{tab:dataset}
\end{table*}

\subsection{Data Capture Methodology} \label{sec:dataset:methodology}

We have written a crawler that automatically collects information on publicly accessible Facebook broadcasts. These are taken from the Facebook Live map,\footnote{fb.com/livemap} which publishes broadcasts in realtime. The crawler collects this data every two minutes, providing a periodic snapshot of all geo-tagged public streams in the system (including their metadata). This was executed between 7-Nov-2016 and 2-Dec-2016, resulting in 3TB of data.\footnote{Note that between 18-Nov and 22-Nov the measurements ceased. This was due to an unavoidable server failure. Consequently, this time period is omitted from our later analysis.} This covered 6.5 million broadcasts by 3.29 million unique broadcasters, and viewed by a peak of 62 million users. Amongst several events captured, our dataset covered the US Presidential election of Nov 2016, Thanksgiving (26th November) and Armistice Day (11th November). 

On Facebook Live, a broadcaster can go live using any of the following three mechanisms:
\one~\textbf{From App}: Users or pages can create live streams using the Facebook App or Facebook Mentions\footnote{https://newsroom.fb.com/news/2015/08/connect-with-public-figures-through-live/} from their Android or iOS devices. Note that the feature of going live from the browser was introduced three months after the crawl.\footnote{https://newsroom.fb.com/news/2017/03/new-ways-to-go-live-now-from-your-computer/}
\two~\textbf{From Publisher tools\footnote{https://www.facebook.com/facebookmedia/get-started/live}}: Facebook Pages\footnote{https://www.facebook.com/business/learn/facebook-page-basics}  can use this feature allowing them to use any external device or software to stream the live content.
\three~\textbf{Using Developer API}: Developers can use the live API\footnote{https://developers.facebook.com/docs/videos/live-video} and embed it in their app to broadcast their content through Facebook. 
 95\%, 4\% and $<$1\% of broadcasts belong to each of these above category respectively. 

For every broadcast, the following data is collected every two minutes:  
\one~\textbf{Broadcaster metadata}: The broadcaster's username and the geo-tag coordinates (\cf \S\ref{sec:dataset:coordinates}). The broadcaster can be either an individual user account or, alternatively, a Facebook Page (typically created for organisations, \eg political parties). In the latter case, we also collect  information about the category of the broadcast page, \eg Media, Sport, Shopping \etc
Geo-tags are added by the user based on their GPS coordinates; if a user does not share location information explicitly, any publicly accessible location information in their profile will be used. \two~\textbf{Viewer information}: The number of live viewers per broadcast, as well as the location coordinates of the viewers (returned by Facebook).

Engagement attributes are also collected, namely the number of likes, shares and comments during the broadcast. After eight months we then revisited any videos that had been archived for later on-demand access, collecting the number of likes, shares, comments and the bitrate (46.1\% of live streams were archived).
Table~\ref{tab:dataset} provides a summary of the dataset across different regions.

\begin{figure}[tb]
 \centering
\includegraphics[width=0.95\linewidth]{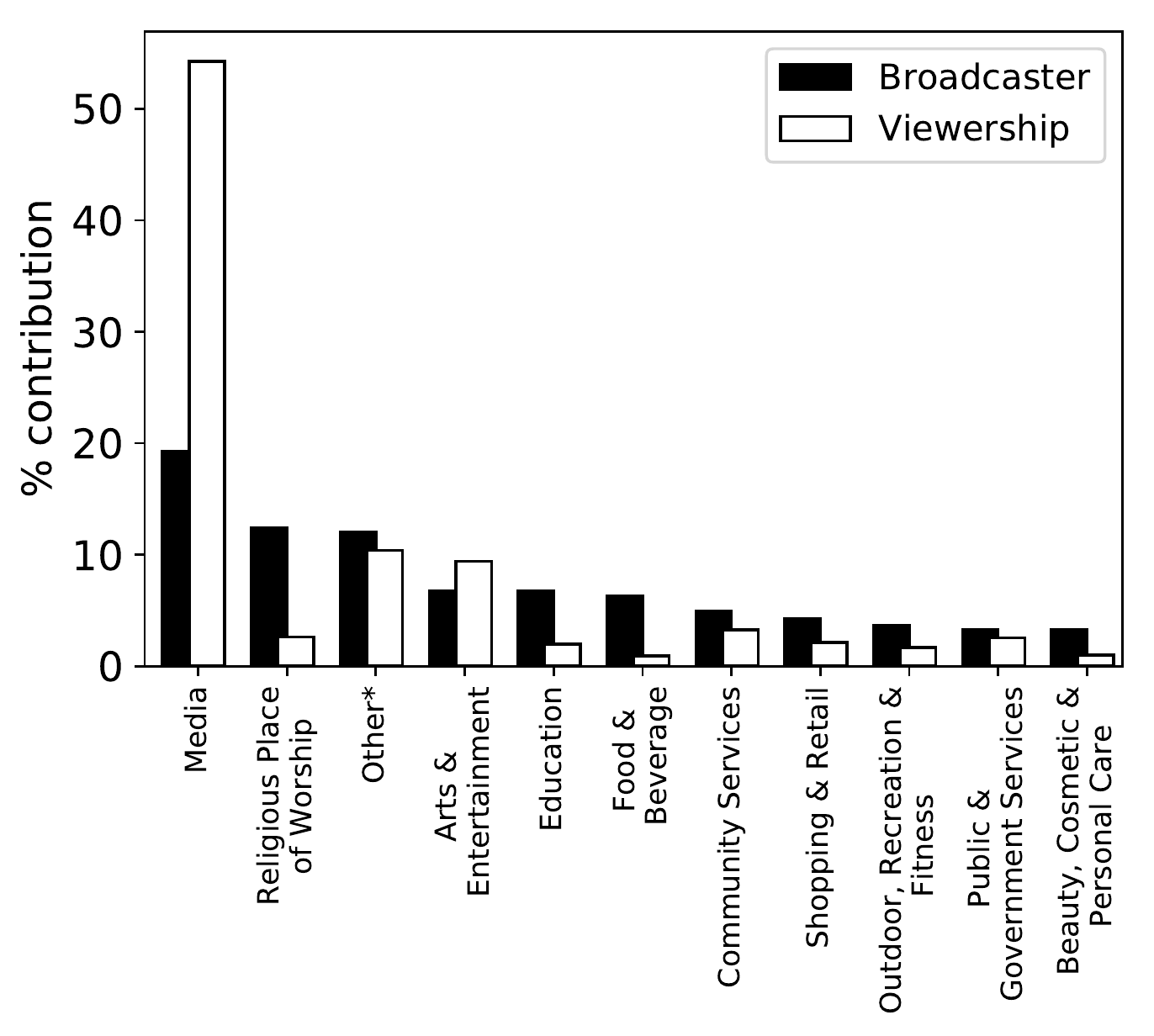}
\caption{Fraction of broadcasters and viewers for each category. Note that ``Other'' is a category provided by the Facebook itself.}
\label{fig:bcaster_vs_vship}
\vspace{-3mm}
\end{figure}

To give context regarding the \emph{types} of content material, we briefly inspect the categorical tags within the Page broadcasts. Figure~\ref{fig:bcaster_vs_vship} presents the fraction of broadcasts that fall into each category, as well as the fraction of the overall viewership that each category attracts. Media, which covers things like news and TV content, gains by far the highest viewership (over 50\%), despite constituting less than 20\% of broadcasts. ``Religious Place of Worship'' is second most popular by the number of broadcasters, driven by the US and Brazil where they constitute over 30\% of the streams. Measuring across the world, however, 40\% of the religious broadcasts do not garner even a single view. Other content types ranging from arts to shopping can be seen of assorted popularities.

\subsection{On Facebook's geo-coordinates}\label{sec:dataset:coordinates}

As stated above, we collect geo-coordinates for viewers and broadcasters. These are used by Facebook to visualise locations on a public map for users. Before continuing,
it is important to validate the fidelity of these coordinates and explain the level of sampling Facebook utilises when returning locations of viewers. Every broadcast is accompanied by a longitude-latitude pair from where the broadcast happens and the viewer locations  are reported for the broadcasts that have greater than 100 views.

\begin{figure}
	\centering
	\includegraphics[width=.9\linewidth]{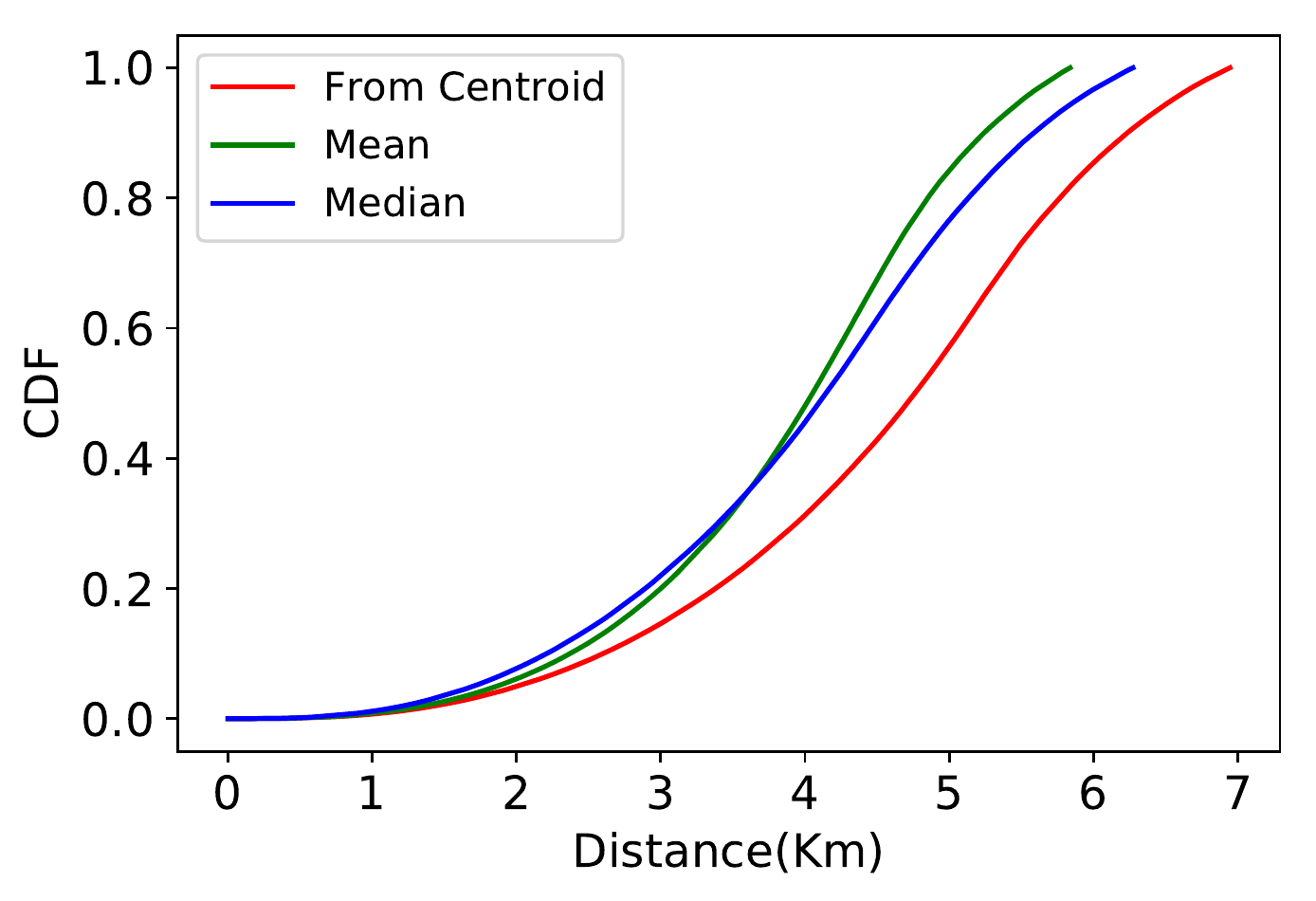}
	\caption{CDF of maximum distance from centroid of locations reported for each broadcaster, as well as the mean and median distance between reported locations per broadcast.}
	\label{fig:latlon_acc}
	\vspace{-3mm}
\end{figure}

\emph{All} the broadcasters have geo-tags, whilst the mean percentage of viewers that also have locations reported is 55\%, with a standard deviation of 12.83. We regularly observe broadcasters reporting multiple locations per broadcast (this could be because the user has moved or because Facebook has introduced noise to protect the location privacy of broadcasters). To explore this, for each broadcast, we take all reported coordinates for the broadcaster and compute the centroid. We then calculate the maximum distance between the centroid and all geo-co-ordinates reported --- this gives the upper bound of any noise potentially introduced. Figure~\ref{fig:latlon_acc} presents a Cumulative Distribution Function (CDF) of the results. We find that \emph{no} broadcaster reports a distance that varies greater than 7 KM from the centroid. This indicates that any noise (if introduced) will only undermine accuracy by this upper bound; this is sufficiently accurate to allow global-scale geo analysis. Finally, we convert each geo-coordinate into its country of origin using country polygons. For subsequent statistics related to distance, we utilise Vincenty's formula for computing the distance between coordinates~\cite{vincenty1975direct}.

\subsection{On Facebook's infrastructure}

Before continuing, we briefly outline how Facebook Live operates from an infrastructural perspective~\cite{fb2015arch}. When users go live, they are redirected to their closest Point of Presence (PoP) using RTMPS. Video data is adaptively encoded by the Facebook mobile app, and then uploaded to a data centre via the PoP. Typically, data is uploaded in a high definition format, but quality is degraded in cases of poor connectivity. Any users wishing to view the stream downloads a manifest file, and begins requesting listed video chunks via Facebook's CDN (mainly Akamai), where caching also takes place.

%% file: sections/char.tex
\section{Characterising Live Broadcasts}\label{sec:char}
We begin by characterising the global activities of broadcasters and their viewerships on Facebook Live.

\subsection{How popular is Facebook Live?}\label{sec:char:pop}

\begin{figure}[t]
 \centering
\includegraphics[width=1.0\linewidth]{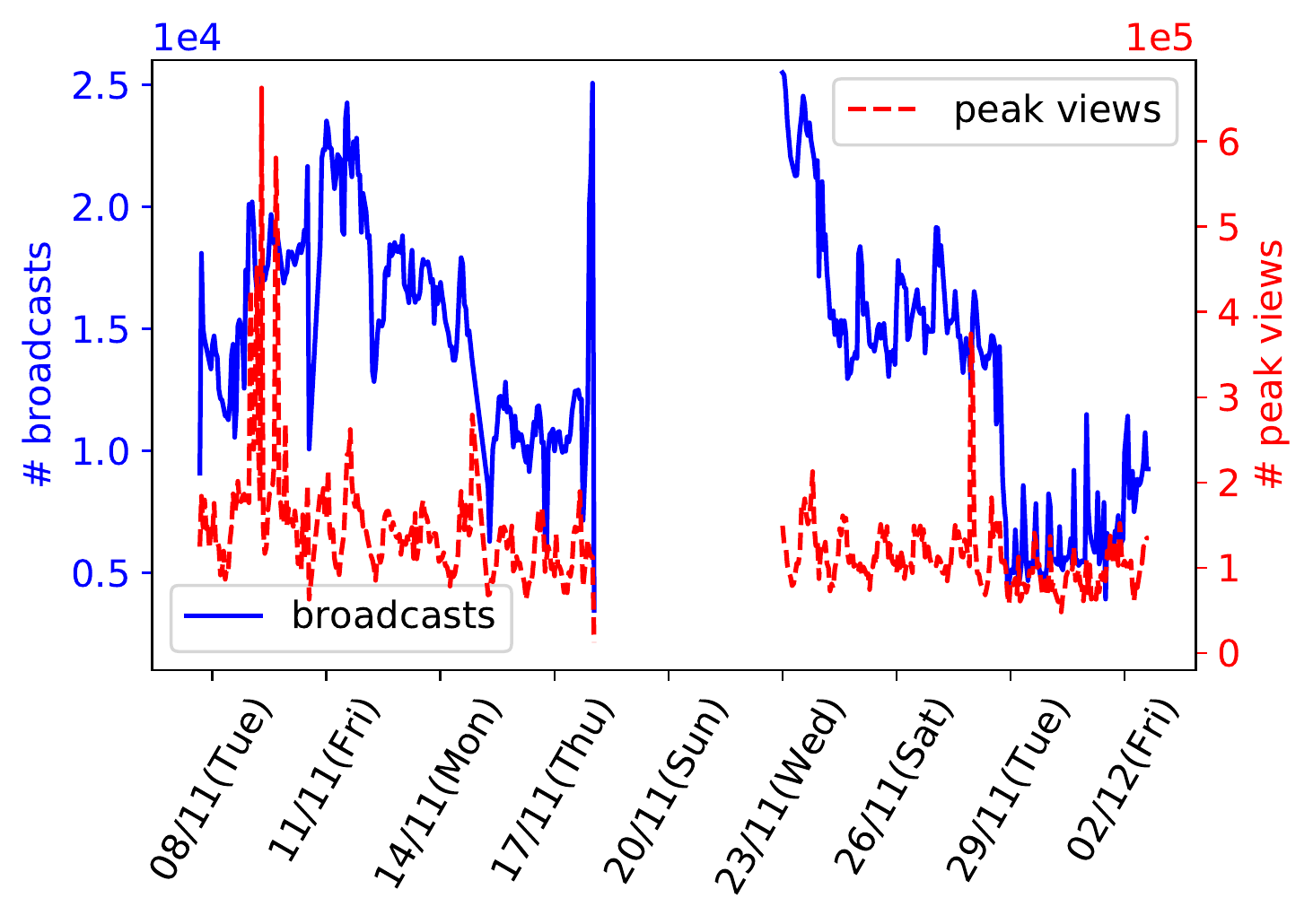}
\caption{Cumulative number of broadcasters and viewers per hour during measurement period. Note that between 18th and 22nd November, monitoring ceased due to a failure in our crawler.}
\label{fig:NC_broadcasts_no}

\end{figure}

First, we look at the overall popularity of the platform, as measured by the number of broadcasters and viewers. Figure~\ref{fig:NC_broadcasts_no} presents an hourly time series across our dataset. It counts both the number of broadcasts (Y1-axis) and viewers (Y2-axis). We see a roughly stable trend for viewers, but erratic trends for broadcasters. This is largely due to a high number of broadcasts in November 2016 due to the US Presidential elections. We also observe periodic spikes in utilisation. These centre on weekends, indicating the social nature of the broadcast platform.

\begin{figure*}[thb]
	\centering
	\subfloat{
		\label{fig:views_pg_vs_user}
		\includegraphics[width=.32\linewidth]{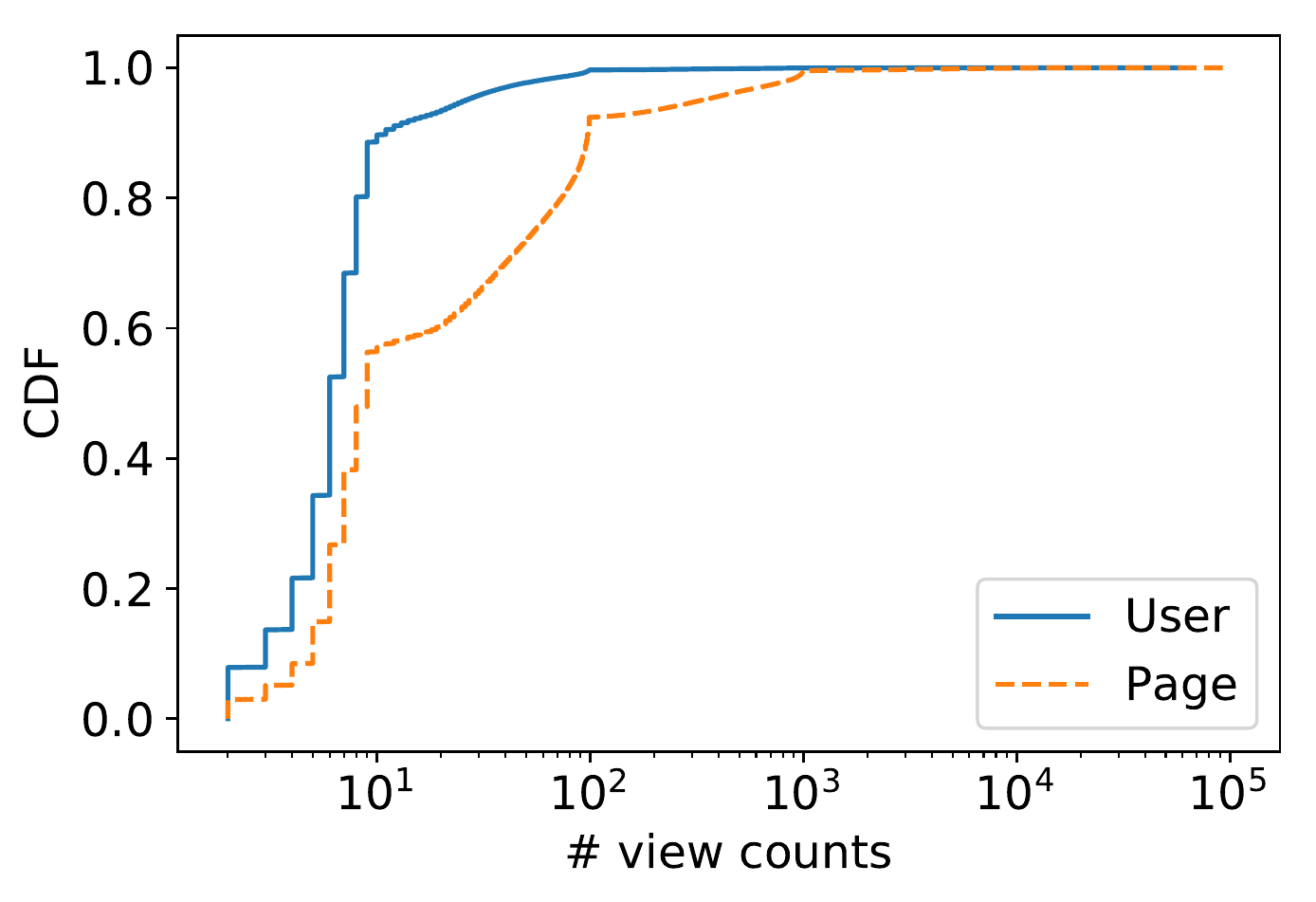}
	}
	\subfloat{
		\label{fig:vid_cat}
		\includegraphics[width=.32\linewidth]{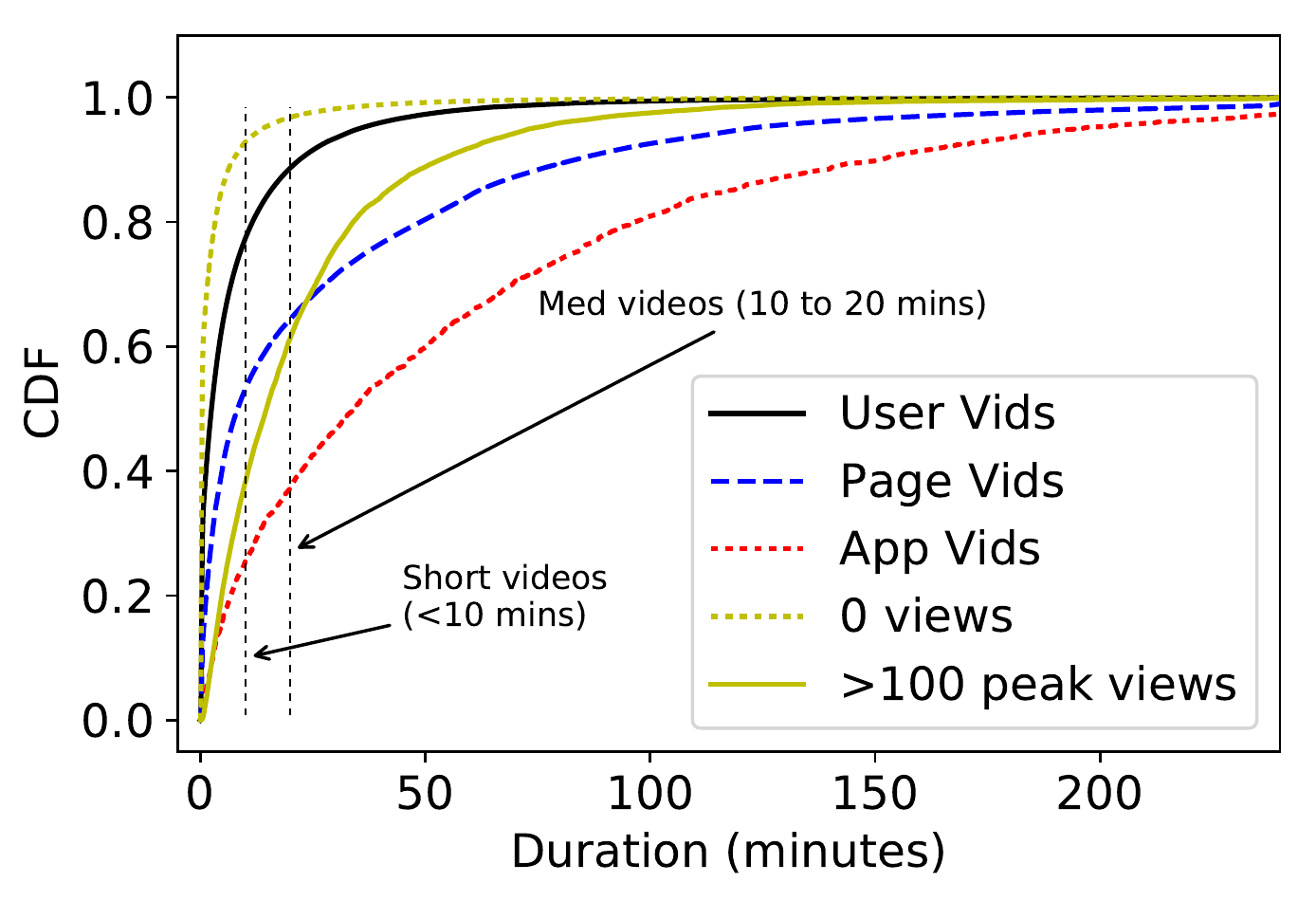}
	}
	\subfloat{
		\includegraphics[width=.32\linewidth]{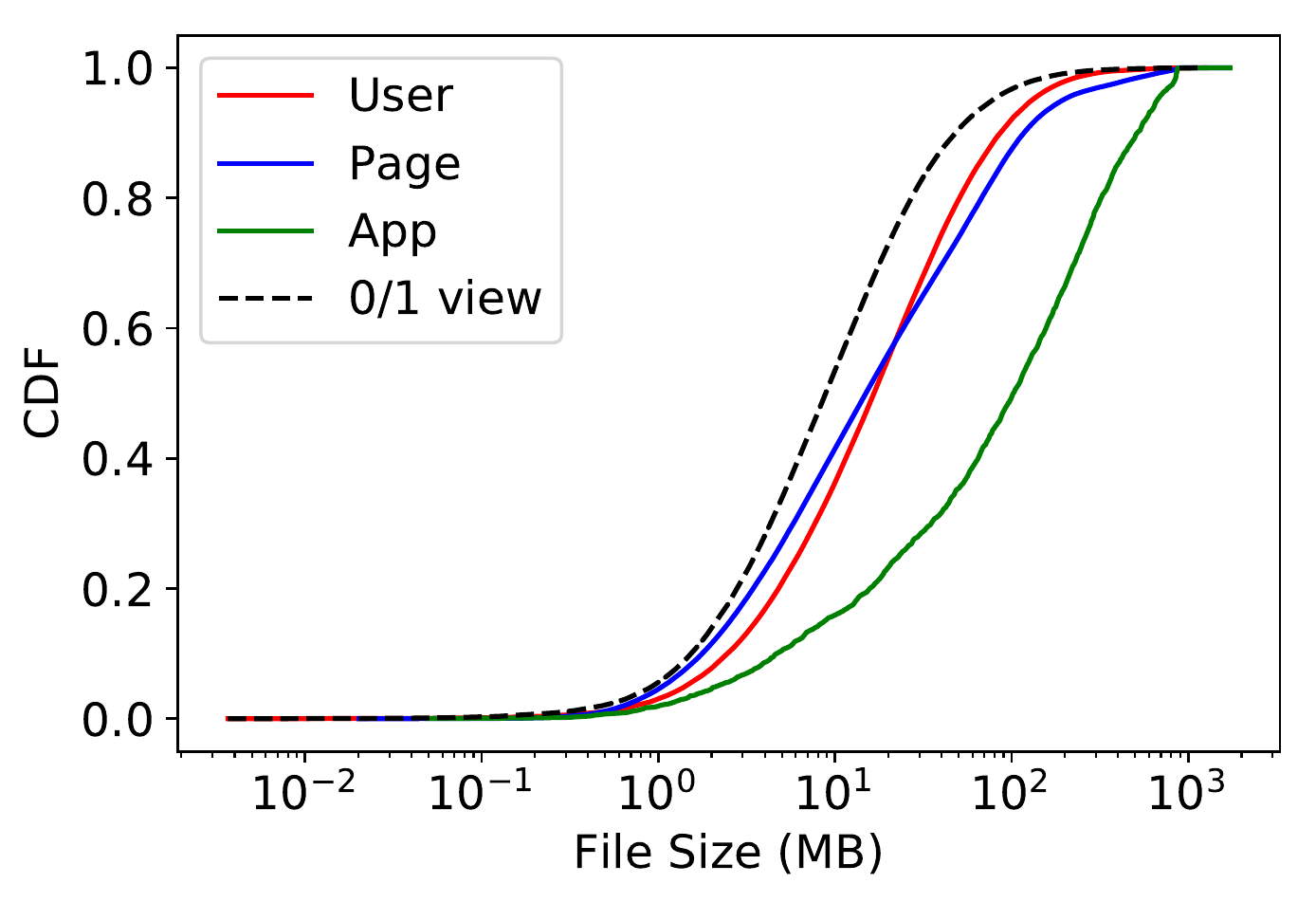}
		\label{fig:file_sizes}
	}
	
	\caption{Broadcast characteristics:
		\protect\subref{fig:views_pg_vs_user} CDF of peak view counts (for broadcasts with a view count > 1). Broadcasts are separated into those generated by Users and Pages. 41.47\% and 37.39\% of the user and page videos go unwatched;   9\% of user videos and 5\% of page videos  get just a single view. \protect\subref{fig:vid_cat} CDF of broadcast duration. Broadcasts are separated into streams generated by Page accounts and User accounts, as well as popular ($>$100 peak views) and unpopular (0 views) streams. \protect\subref{fig:file_sizes} Size of uploaded broadcast files, for user, page and app videos, as well as  videos with 0 or 1 viewers. 	
	}
\vspace{-3mm}
\end{figure*}

Whereas the above quantifies how popular Facebook Live is, it does not demonstrate the  popularity of individual broadcasters. This is typically a more important metric for users wishing to use their broadcasts for promotion. Figure~\ref{fig:views_pg_vs_user} presents a CDF of the peak view counts\footnote{Note that the number of viewers changes as viewers join and leave a broadcast. Since our trace collects data at 2-minute intervals, we show the peak number of viewers recorded.} for each broadcast. Two lines are present: \one~Broadcasts by \textbf{User} accounts, which are generated by individual Facebook users on their mobile device and \two~Broadcasts by \textbf{Page} accounts, which are generated by Facebook Pages. 

It can be seen that Page broadcasts gain significantly more viewers than typical user broadcasts. To add further context, Table~\ref{tab:top_broadcasters} presents the Top 10 broadcasters, ranked by peak viewing figures. It can be seen that 7 out of the Top 10 broadcasters are Pages. 
Even though Page broadcasters only contribute 4.26\% of the streams, they collect almost 35\% of all views.
In stark contrast, we find that 41.5\% of user broadcasts \emph{never} gain a single viewer, whilst 55.35\% of broadcasters have at least one stream that remains unwatched. In fact, the median view count for user broadcasts is just 1. This raises questions regarding the extent that Facebook Live is actually used as a social \emph{broad}cast medium in practice, as opposed to a unicast stream that is seen by one or even zero viewers.

\begin{figure}[th]
	\centering
	\includegraphics[width=1.0\linewidth]{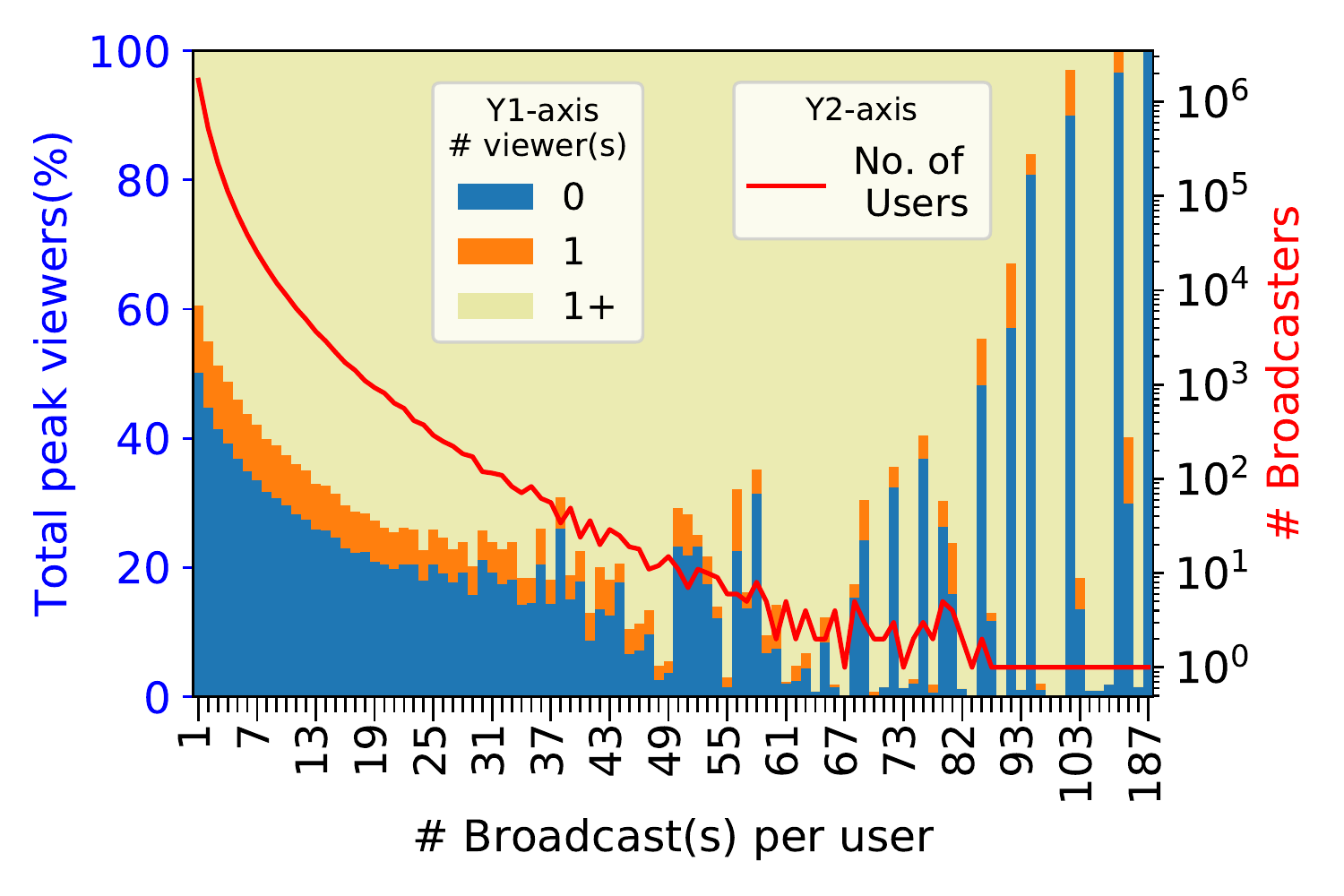}
	\caption{Distribution of number of peak viewers (Y1-axis on the left) bucketed based on the broadcast(s) made by an user. Y2-axis (on the right) shows the total number of broadcasters falling under a bucket of total broadcasts performed.}
	\label{fig:bcast_per_user}
	\vspace{-3mm}
\end{figure}

To study this further, we separate all users into buckets based on the total number of broadcasts they have performed. For each bucket, we compute the fraction of broadcasts that gained  0 viewers, 1 viewer and $>$1 viewers. Figure~\ref{fig:bcast_per_user} presents our findings as a stacked bar chart; buckets are plotted on the X-axis and the fraction of broadcasts with zero views is plotted on the Y-axis. We also plot the number of samples in each bucket on the Y2-axis. Unsurprisingly we find on the Y2-axis that the number of broadcasters in each bucket decreases as the threshold for number of broadcasts increases. Focusing on the Y1-axis, we find, as expected, that the heavy users who broadcast more have a lower proportion of unwatched streams. However, we observe that \emph{all} user groups generate unwatched material. This extends to users who generate tens of broadcasts. This confirms that unwatched material is not simply created by experimental users who just ``try out'' Facebook Live once. Instead, it is a persistent characteristic of broadcasters. For example, for users with 10 broadcasts, we find that 29.56\% of broadcasts gain 0 viewers, 7.9\% gain 1 viewer and the remainder gain above 1 viewer. Interestingly, the graph also demonstrates a valley-like trend, whereby users with a small number of broadcasts largely gain 0 viewers (45.8\% for below 5 broadcasts), whilst users with a high number of broadcasters also generate a lot of unwatched streams (36.35\% for above 100 broadcasts). Users who fall in the middle of the spectrum (40--50 broadcasts) have the lowest proportion of unwatched material. Although this observation is impacted by the smaller sample sizes for user groups with large numbers of broadcasts (as shown on the Y2-axis), our manual inspection reveals that such ``unpopular'' users tend to generate large numbers of uninteresting material, \eg unprocessed personal diaries. It should also be noted that the generation of unwatched material is not exclusive to mobile users; some categories of Page streams exhibit similar properties, \eg 40\% of ``Religious'' streams from Page accounts are never viewed, and 13\% for ``Media'' streams. Overall, 37.4\% of Page streams have zero views.

\begin{table*}[th]
	\centering
	\caption{Top broadcasters based on median view count.}
	\begin{tabular}{|l | l | S[table-format=5.0,group-digits=integer,group-four-digits] | l | l | l | l |}
		\hline
		Handle & Type & \text{Median Viewers} & Description  & Category & Verified \\
		\hline\hline
		@zuck & User & 48471 & Facebook founder  & User & Yes\\
		@GilmoreGirls & Page & 39162 & Popular TV drama & TV Programme & Yes\\
		@buzzfeedtasty & Page & 34369  & Food news outlet & Media & Yes \\
		@pena6789  & User & 16952 & Adult content & User & NA \\
		@WhiteHouse & Page & 16124 & Official WH page & Historical Landmark  & Yes\\
		@FoxNews  & Page & 14721 & News outlet & Media & Yes\\
		@bendeac & User & 13289 & Romanian Actor & User & No \\
		@RealMadrid & Page & 12783 & Popular football team & Sports Team & Yes\\
		@WesternJournalism & Page &  10507 & News outlet & Media & Yes \\
		@BuzzFeed & Page & 10276 & News outlet & Media & Yes\\
		\hline
	\end{tabular}
	\label{tab:top_broadcasters}
\end{table*}

\subsection{How long are broadcasts?}\label{sec:char:duration}

The previous section has highlighted that nearly half of all user broadcasts are never viewed. We next inspect the durations of these broadcasts to understand the resources consumed (both in terms of user time and upload volume). Figure~\ref{fig:vid_cat} presents a CDF of the duration of broadcasts separated into a number of groups. App broadcasts are, by far, the longest with 53.46\% exceeding half an hour. This is largely driven by the fact that most apps broadcast computer gameplay. Broadcasts performed by Pages also tend to be substantially longer in duration than User broadcasts. Almost 30\% of the Page broadcasts exceed half an hour, compared to just 6\% for User broadcasts (where 77.34\% are under 10 minutes).

This trend may be partly driven by the greater popularity of Page streams (\cf Figure~\ref{fig:views_pg_vs_user}). To examine this more closely for mobile users, Figure~\ref{fig:views_pg_vs_user} breaks User broadcasts into two groups: \one~\textbf{Popular}, which gain over 100 viewers at peak; and \two~\textbf{Unpopular}, which are never viewed. In-line with our previous statement, we see that more popular streams broadcast for longer. 25\% of these streams last longer than half an hour, whereas this is just 2\% for unpopular streams. Critically, however, we still find a non-negligible number of unwatched streams that broadcast for extended periods (80k videos are longer than 20 minutes). 

Currently, streams with zero views are live streamed to Facebook's infrastructure regardless of their popularity. 
This observation suggests significant waste in network, client and server resouces. For example, the average duration for unwatched streams is five and a half minutes. With a bitrate of 500kbps, this would generate 18MB of needless traffic per broadcast (with battery and network costs). This therefore raises the question of how such content is delivered. Whereas uploads are obviously necessary for archived streams (\ie ones that remain on a user's profile), this is profligate for live streams that are not archived. This is because such data is immediately discarded without anybody ever seeing it. Importantly, we find that only 41.7\% of unwatched streams are actually archived --- for the remainder, it is unnecessary to waste device resources in performing the live upload to the PoP.  This observation leads us to inspect the size of the archived videos (\ie videos marked by the users as being available for on-demand viewing later on) to understand the precise volume of this upload process. Figure~\ref{fig:file_sizes} presents the distribution of file sizes for all archived videos. The median file size for broadcasts with zero views is 8.57 MB, with 45\% exceeding 10 MB.

\subsection{Is broadcast really necessary?}
\label{sec:char:necessary}

Combining the above observations, we see that Facebook Live is predominantly (95\%) used by mobile users to broadcast content. Pages garner a disproportionate number of viewers, whilst 41.47\% of user broadcasts languish in obscurity, never being viewed. These observations indicate that the current ``cloud-based'' model of delivery is misplaced. As it stands, the Facebook mobile app uploads content regardless of view counts. This opens up interesting opportunities for hyper-local storage. For example, video content can be cached locally on the mobile device, and only uploaded once the first viewer arrives\footnote{To entice new viewers, a short GIF or video highlight of a few seconds can be shown, to indicate the ongoing video broadcast.}.
Upon the completion of a broadcast, users are given the option of archiving the content.
Live content that is not watched and not archived can be cached locally and then discarded after broadcast. If the user wishes to archive the video, it can be automatically uploaded to Facebook when convenient (\eg via WiFi).
This simple change would reduce use of battery and network for the 55.58\% of broadcasters who (at least once) generate an unwatched video. It would also save an average of 5.84MB per upload. Note that this also covers broadcasters who generate \emph{any} video chunks that are never watched. For example, 44.1\% of streams in their first 2 minutes have zero viewers, out of which 7.8\% continue to garner viewers later. Hence, such users need not upload the first chunks until a viewer arrives\footnote{In the current Facebook live design, new viewers cannot rewind on a live stream; thus these first chunks need not be uploaded even after the first viewers arrive.}.

%% file: sections/geo.tex
\section{Geographical Exploration}\label{sec:geo}

The previous section has highlighted the nature of social broadcast, focussing on the large number of unwatched and unpopular streams. Next, we inspect the opposite end of the spectrum --- highly popular streams. A unique aspect of Facebook Live is that it provides the locations of stream viewers for broadcasts with an audience of 100+. This allows us to understand the geographical relationship between broadcasters and viewers.

\subsection{Where are the users?} \label{sec:geo:where}

\begin{figure}[t]
	\centering
	\includegraphics[width=1.0\linewidth]{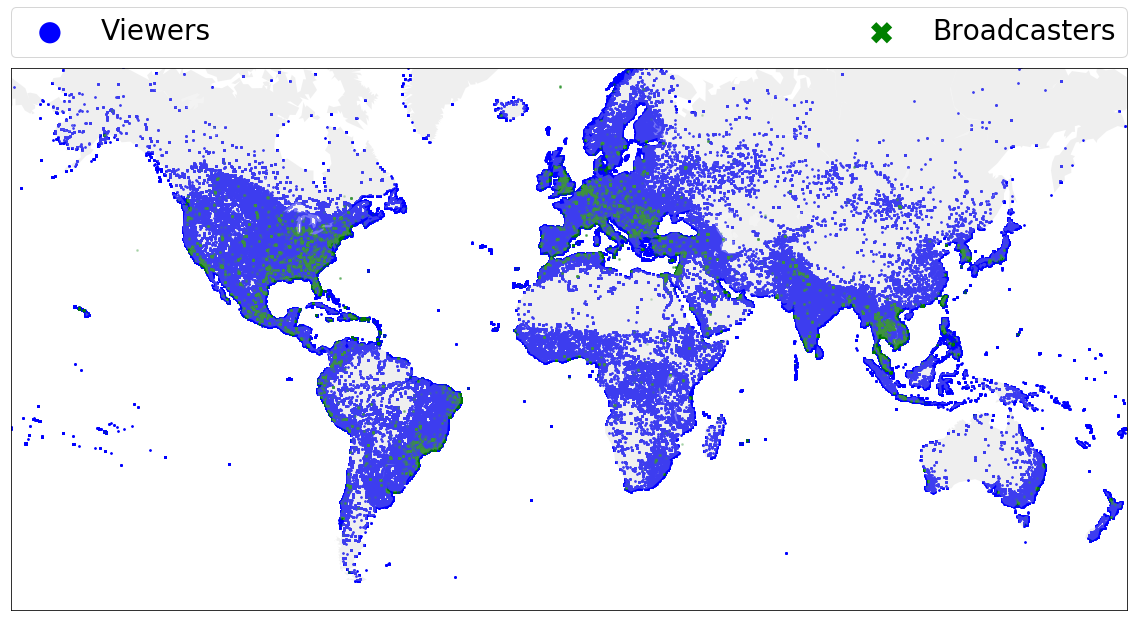}
	\caption{Locations of influential broadcasters ($>$100 viewers) and the respective viewers across the globe}
	\label{fig:bcast_view_map}
	\vspace{-2mm}
\end{figure}

We begin by  inspecting the locations of users to understand where Facebook Live's userbase is. Figure~\ref{fig:bcast_view_map} presents viewer (blue) and broadcaster (green) locations for any streams that garner above 100 viewers. This covers 45K broadcasts, and 20.3 million viewers. The map shows a significant spread of broadcasters (green) with a focus on East Coast US, Brazil, Europe and South East Asia. It can be seen that most major conurbations are covered. A particularly nice feature of the dataset is that locations are based on mobile GPS-based geo-tags (rather than IP address). This gives us accurate geographical vantage into regions such as China, which typically access Facebook via VPN (thereby changing their reported IP geolocation). It can be seen that despite the blocking of Facebook, there are a number of users in China. In the rest of this section, we focus on the top 19 countries measured by broadcasters and consumers. These countries are obtained taking the union of the top 15 countries in terms of broadcasters and top 15 in terms of consumers. \looseness=-1

\begin{figure}[h]
	\centering
	\includegraphics[width=0.93	\linewidth]{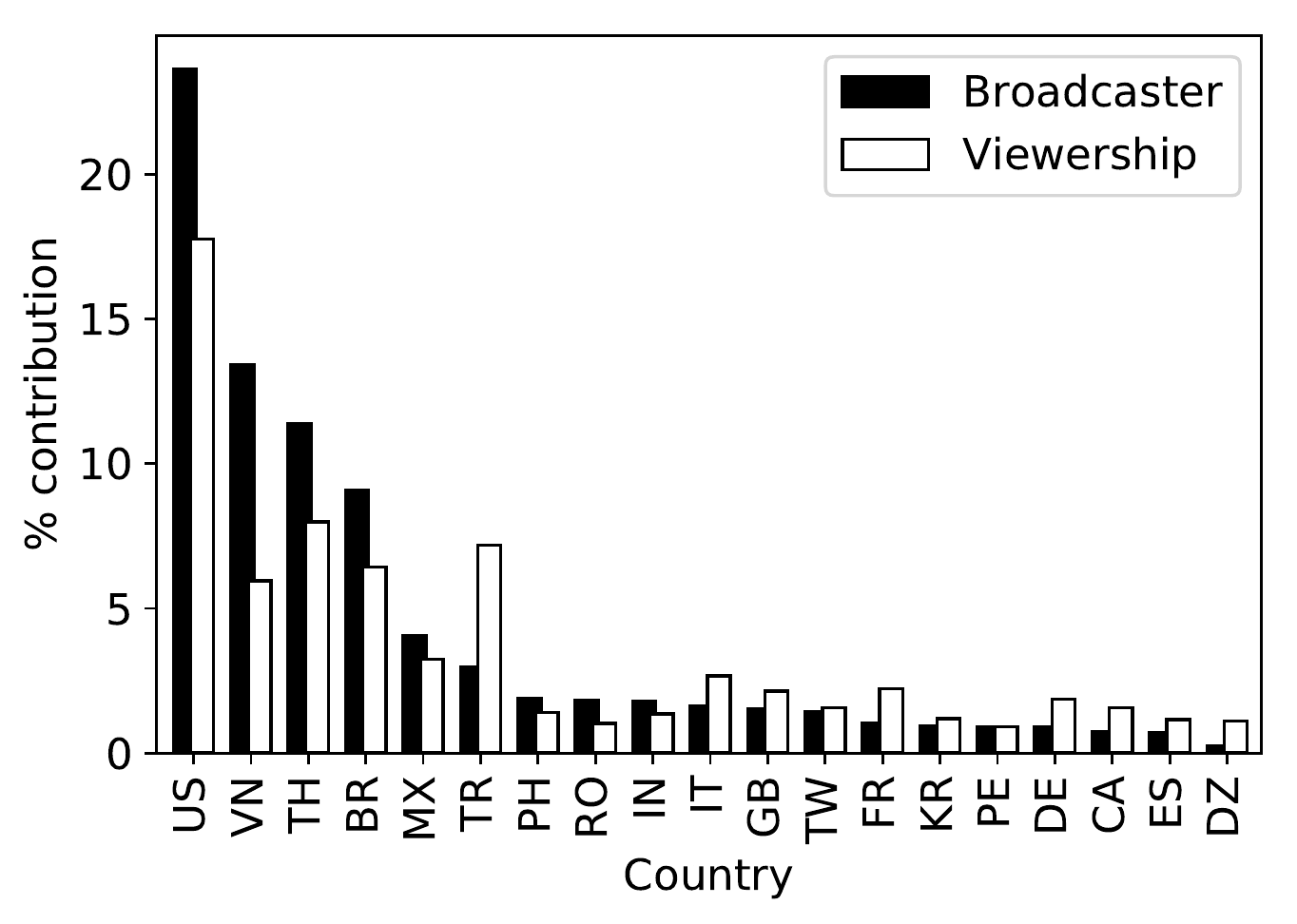}
	\caption{Contribution of broadcasters and viewers from top-accessed countries.}
	\label{fig:bcaster_vs_viewer}
\end{figure}

Figure~\ref{fig:bcaster_vs_viewer} presents the percentage of broadcasters and viewers based in each of the top 19 countries. A clear ranking can be observed; the US dominates both the number of broadcasters and viewers, followed by Thailand and Vietnam. The latter is particularly interesting, as in many cases we find that the number of viewers and broadcasters in a country do not necessarily correlate. For instance, whereas, in the US or Brazil, the fraction of viewers and broadcasters is relatively \emph{similar}, Thailand generates a significant proportion of the world's broadcasts (11.26\%), but only constitutes 8\% of the viewers. Conversely, in Turkey, there are significant fraction of viewers (7.1\% of the world's viewers), but only 2.94\% of broadcasters. Similar observations can be made across many other countries, including Korea, Peru and Romania. Through manual inspection, we find a number of potential reasons. For instance, Turkey blocks access to Facebook; therefore, most users are likely accessing it via circumvention tools. Whereas Turkish users may be keen on consuming content, they may be less willing to expose themselves by broadcasting on a censored medium. In the case of Thailand, we find that a notable portion of adult content is being streamed, and this niche content attracts a globally distributed viewership (rather than just in Thailand).

\subsection{How far away are viewers?}

\begin{figure*}[thb]
 \centering
 \subfloat{
 	\label{fig:vship_dist_cwise}
 	\includegraphics[width=.34\linewidth]{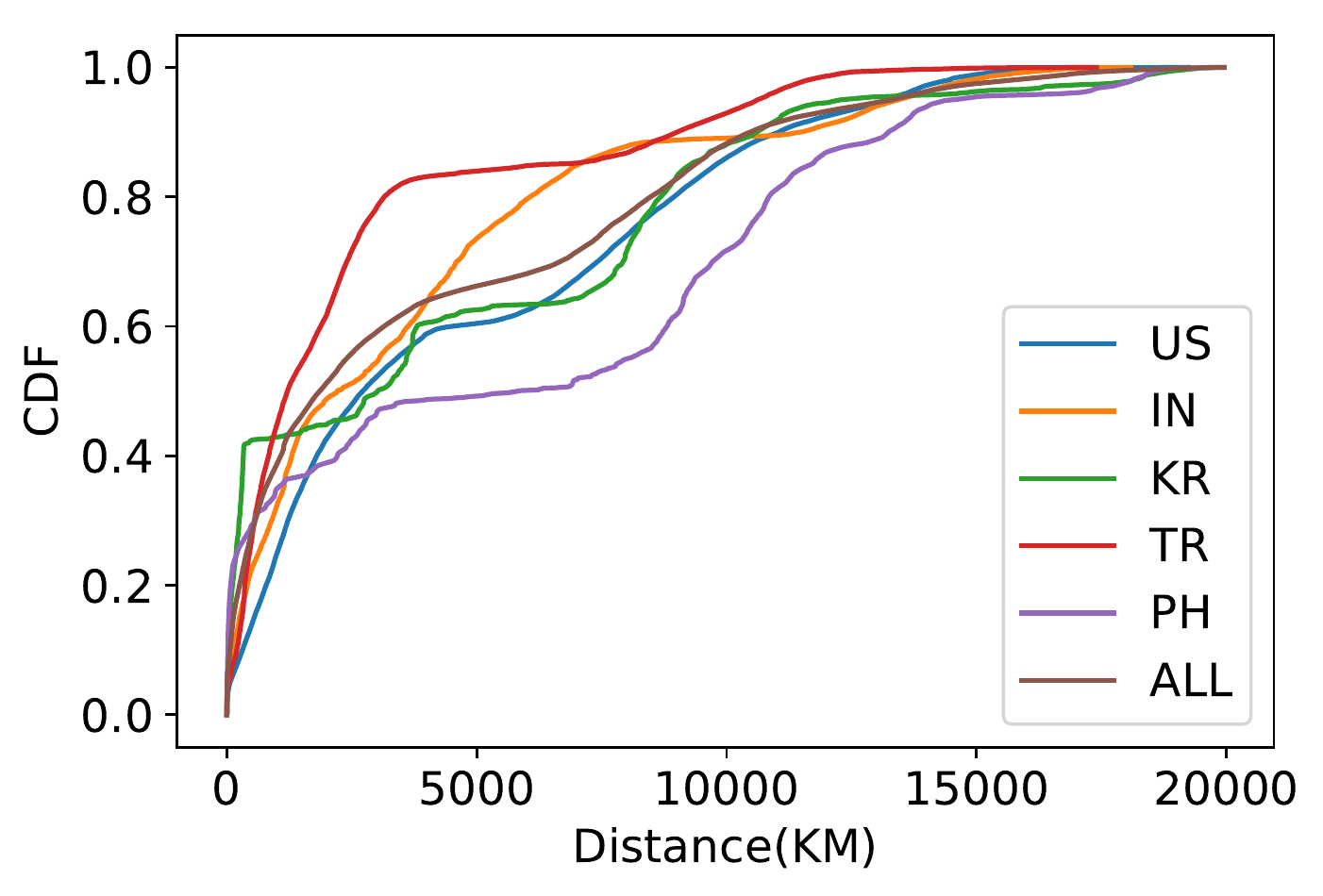}
	}\hfill
 \subfloat{
	 \includegraphics[width=.34\linewidth]{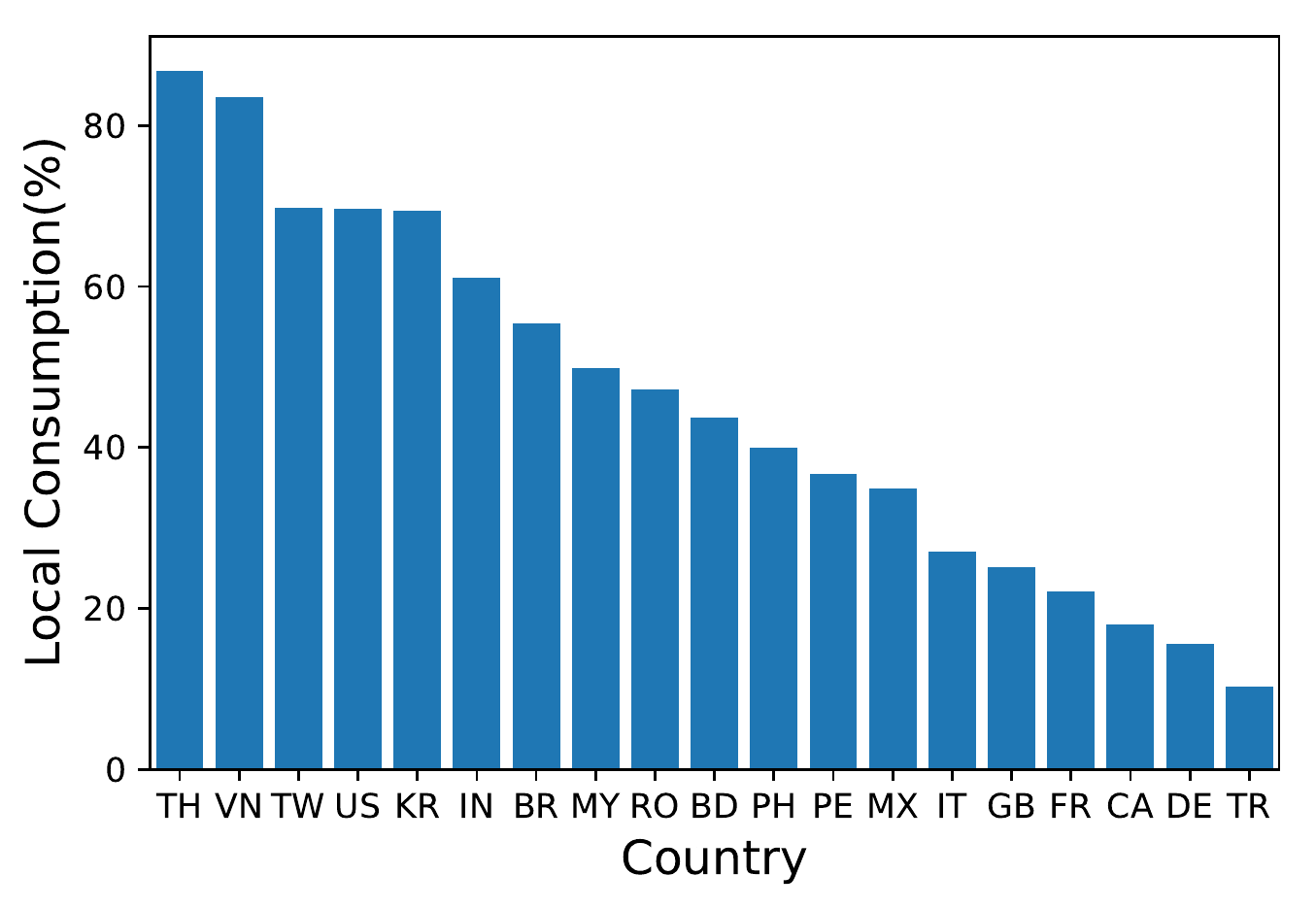}
	 \label{fig:mob_consumption}
	}\hfill
 \subfloat{
	\includegraphics[width=.29\linewidth]{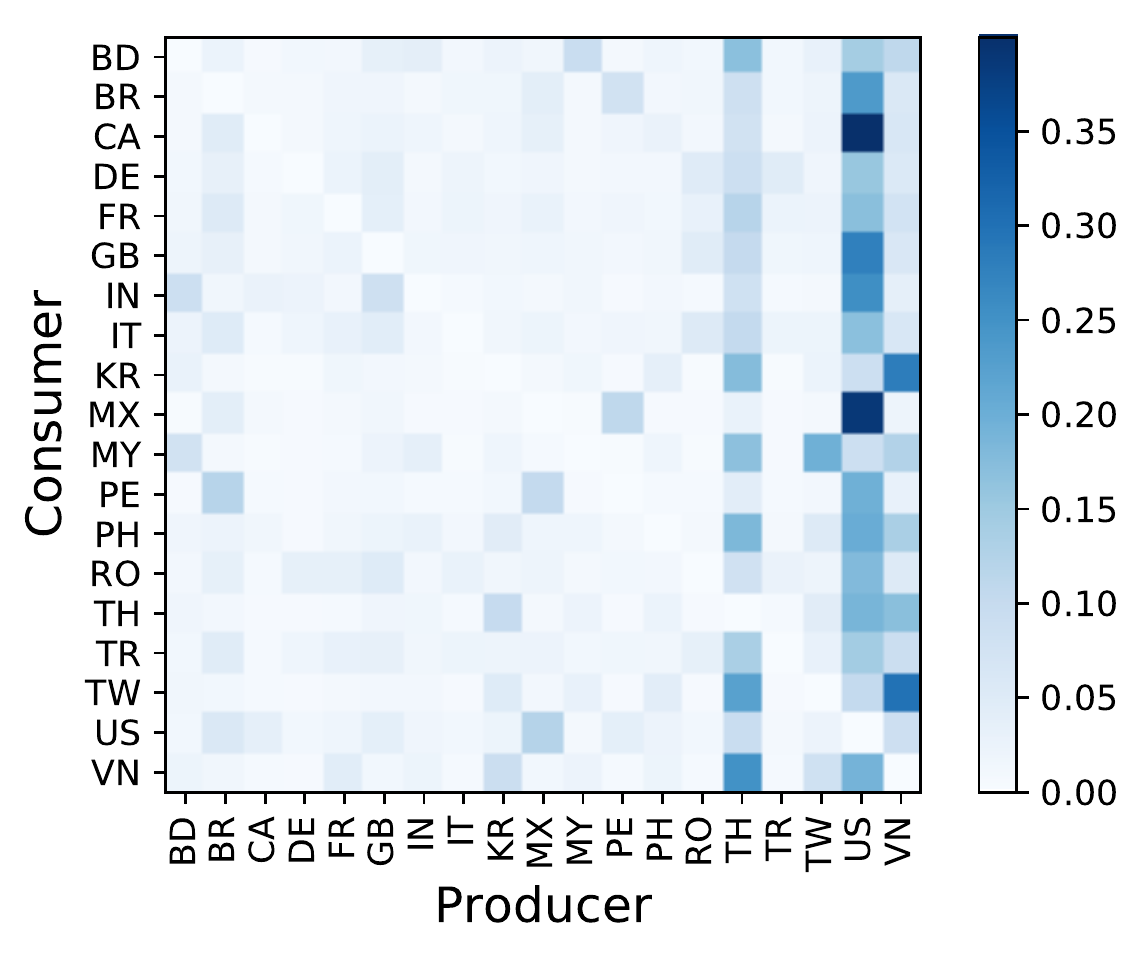}
	\label{fig:coun_heatmap}
	}
\caption{Geographical footprint: \protect\subref{fig:vship_dist_cwise} CDF of distance between broadcasters of their viewers. Multiple plots are presented for broadcasters in different countries
\protect\subref{fig:mob_consumption} Fraction of views from users in the same country as the broadcaster
\protect\subref{fig:coun_heatmap} Fraction of views from users in a \emph{different} country from the broadcaster, shown as a heat map of content broadcast from one country consumed in another country, plotted for the union of top 15 producers and consumers.
}
\vspace{-2mm}
\end{figure*}

Although the above reveals the locations of broadcasters and viewers, it does not show their relationships. From a systems-perspective, this is critical for optimising delay-\emph{in}tolerant communications as ideally viewers, broadcasters and any intermediate servers would be within short distances. Regarding the latter, there have been a number of recent proposals regarding the placement of servers within cell towers and local network PoPs, \ie so-called mobile edge computing~\cite{beck2014mobile, wang2017survey, mach2017mobile}. Thus, we next inspect the distance that viewers are from the broadcasters they watch. Figure~\ref{fig:vship_dist_cwise} presents CDFs of the distance between broadcasters and their viewers for user accounts. We separate broadcasters into their respective countries. In cases where the distribution plateaus, typically the broadcasters and viewers are separated by an ocean (\eg in the case of Philippines this is the distance across the Pacific to reach US-based viewers, who constitute 9\% of the country's audience).

\begin{figure}[t]
	\centering
	\includegraphics[width=0.9\linewidth]{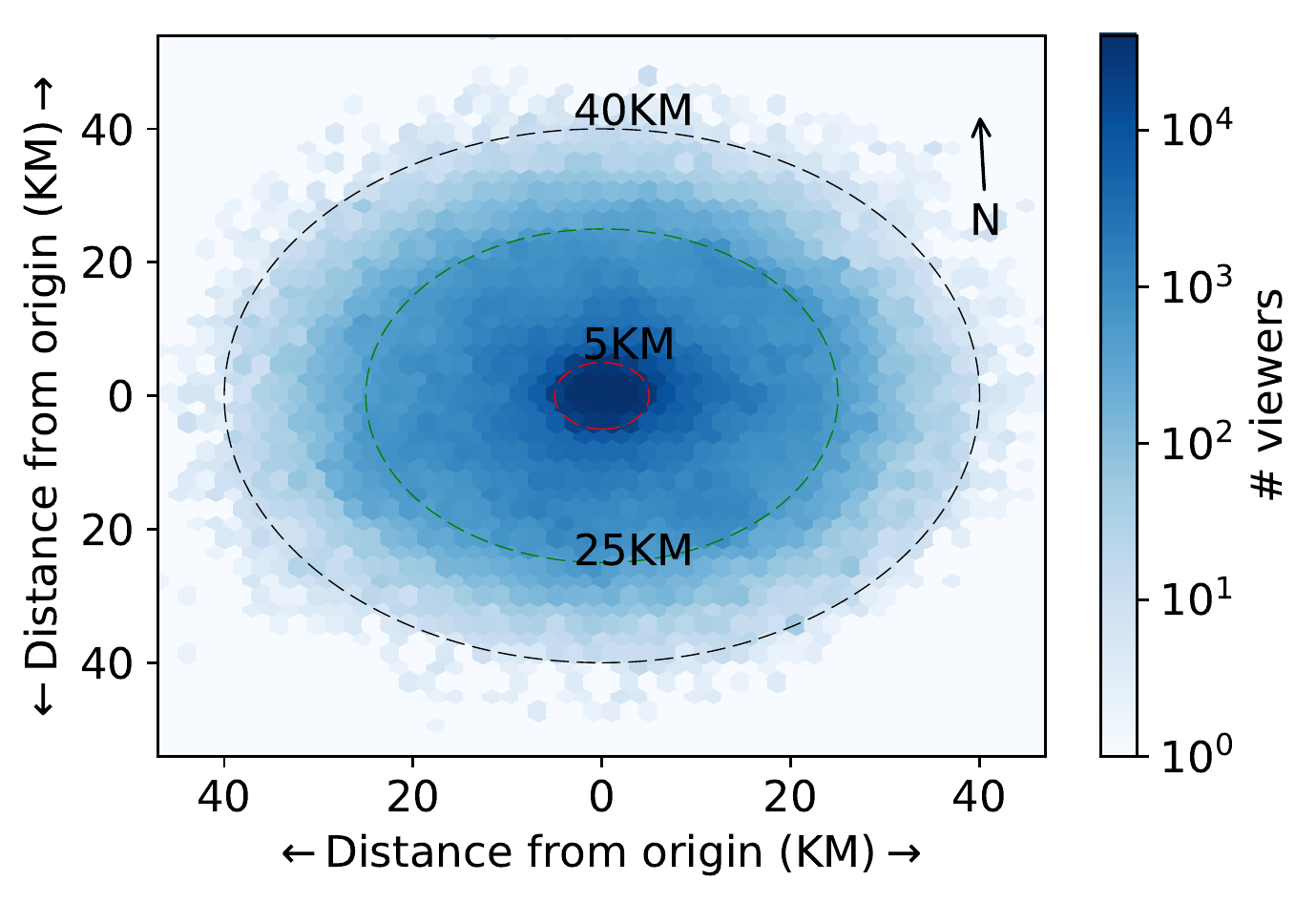}
	\caption{Number of viewers (binned at 500m) within a radius of 5KM, 25KM and 40KM when all the broadcasters location are shifted to origin.}
	\label{fig:local_consumers}
	\vspace{-2mm}
\end{figure}

It can be seen that a notable subset of broadcasters are hyper-local to their consumers. 
This is shown well in Figure~\ref{fig:local_consumers}, which presents a heatmap of viewer locations relative to the broadcasater. 
Overall, 8\% of viewers are within 25 KM of the broadcaster and constitute a city-level locality. South Korea is the most extreme in this regard; 11\% of viewers are within 25KM, with 4\% actually under 5KM. We can also inspect the opposite end of the spectrum. 11.35\% of viewers are over 10,000 KM away from their broadcaster. Here, subtle differences can be seen between broadcasters from different countries. For example, broadcasters from the Philippines, a tiny island in the middle of Western Pacific, tend to be further away from their viewers (only 37.81\% of viewers are closer than 1000 KM), whereas Turkish broadcasters generally garner more nearby viewers (52.71\% are closer than 1000 KM). As previously discussed, this local content consumption suggests that live social streaming would be well placed to benefit from technologies such as mobile edge computing.

\subsection{Domestic or international?}

It should be remembered that distance can be deceptive due to the various sizes of individual countries. Since we expect infrastructure to be confined within country borders, we next separate all broadcasters and viewers into their respective countries. Figure~\ref{fig:mob_consumption} presents the fraction of the views for videos of a country which come from domestic viewers. This is computed as follows: For a given country, we take the total view count of all videos broadcasted from that country. We then present the \emph{fraction} of these views that are domestic, \ie from viewers in the same country as the broadcaster.  In some cases (\eg Thailand, Vietnam), over 80\% of views are  domestic, suggesting strong language, social and cultural ties drive consumption. In other cases, under 20\% of viewers are domestic (\eg Germany). This suggests such countries generate highly popular international content. However, we note that almost all videos have a non-trivial local viewership; and thus would benefit from caching locally in the country of broadcast, regardless of whether the video also has an international appeal or not.

To explore the international views, Figure~\ref{fig:coun_heatmap} presents a heat map of the \emph{non-domestic} views. Each cell in the matrix represents the fraction of views for videos of a country which come from a specific consumer country. For the sake of clarity, the fraction of domestic views (where the producer and consumer are from the same country) are not plotted. Further, only the top 19 producers and consumers are shown; thus the fraction of the heat map values across all consumers might not add up to 1. 

It can immediately be seen that a small number of countries stand-out in regards to the global reach of their content. The US is the most prominent, with a particularly large build-up of viewers in Mexico and Canada. This is not surprising considering the high degrees of immigration and cultural overlap between these three countries. Despite our prior observation about local consumption, Thailand and Vietnam also have a large number of viewers from other countries. Much like the US, this is driven by the large number of broadcasts in the country.  
   
Before concluding, we note that Figure~\ref{fig:coun_heatmap} is intended to illustrate some interesting international trends and is not a complete picture -- for instance, countries like Germany and Turkey, which have a high proportion of non-domestic views (Figure~\ref{fig:mob_consumption}) do not show up with high values in this heat map because the viewers for the video broadcasts from these countries are not amongst the top 15 consumers.\looseness=-1

%% file: sections/social.tex
\section{Understanding Engagement} \label{sec:social}

Our final analysis briefly explores engagement factors. We explore the extent to which views, and other social engagement measures such as numbers of likes, comments and shares evolve over the lifetime of a video, both during the live broadcast, as well as afterwards, and draw implications for mobile livecasting. The main takeaway is that most of the engagement happens \emph{after} the live broadcast, lending further support to the idea that Facebook Live can be treated similarly to an on-demand service, as opposed to a live broadcast.\looseness=-1

\subsection{Evolution of views over time}
First, we capture the instantaneous number of viewers for \emph{all} broadcasts. We do this until minute 34 ($\approx$95\% of videos end before 34 minutes). The number of viewers depends to some extent on the quality of the video, but also upon the duration -- longer videos have a higher chance of catching friends who come online, and notice the live broadcast. Yet, many of them may lose interest and leave after some time, so the later minutes of the video may lack sufficient viewers. 
To avoid introducing bias through the varying sample size across each time interval bucket, we randomly pick 100k videos for each time intervals, and study the distribution of viewer counts. \looseness=-1

\begin{figure}[htbp]
	\centering
	\includegraphics[width=1.0\linewidth]{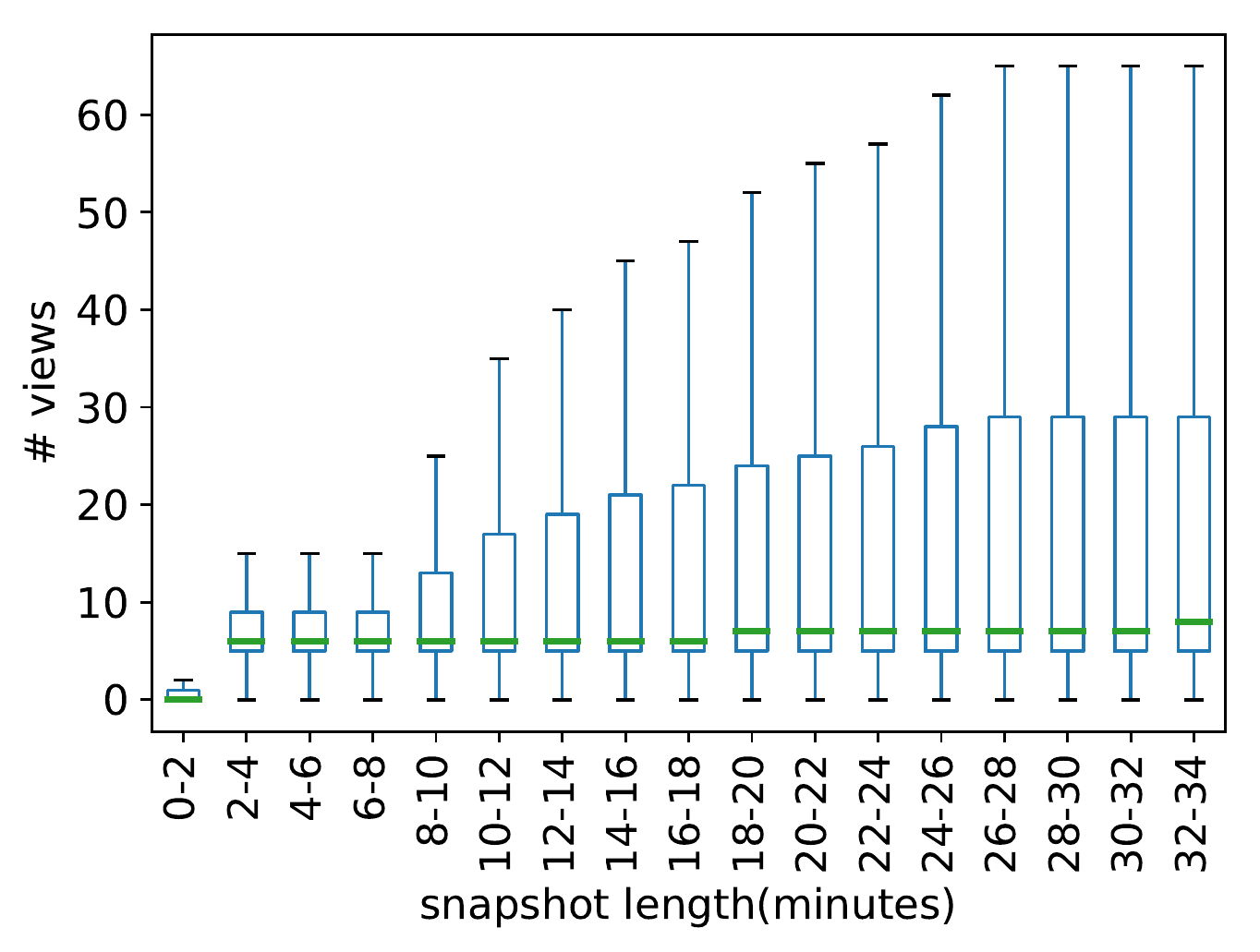}
	\caption{View evolution over 2 minute intervals }
	\label{fig:two_min_view}
	\vspace{-2mm}
\end{figure}

Figure~\ref{fig:two_min_view} shows a box-and-whiskers plot of the number of viewers in each interval across the videos. Each box extends from the lower to upper quartile values of the number of viewers, with a line at the median. The whiskers extend from the box to show the range of the data. Flier points past the end of the whiskers are not shown for clarity. The viewer counts increase after the first interval, and there is a larger variance in the later time buckets, indicating that sufficiently interesting longer videos accumulate viewers over time. But the median number of viewers remains comfortably under 10, even after 20--30 minutes of broadcast, again underscoring that it is not very difficult to support most live broadcasts.

\begin{figure}[t]
	\centering
	\includegraphics[width=1.0\linewidth]{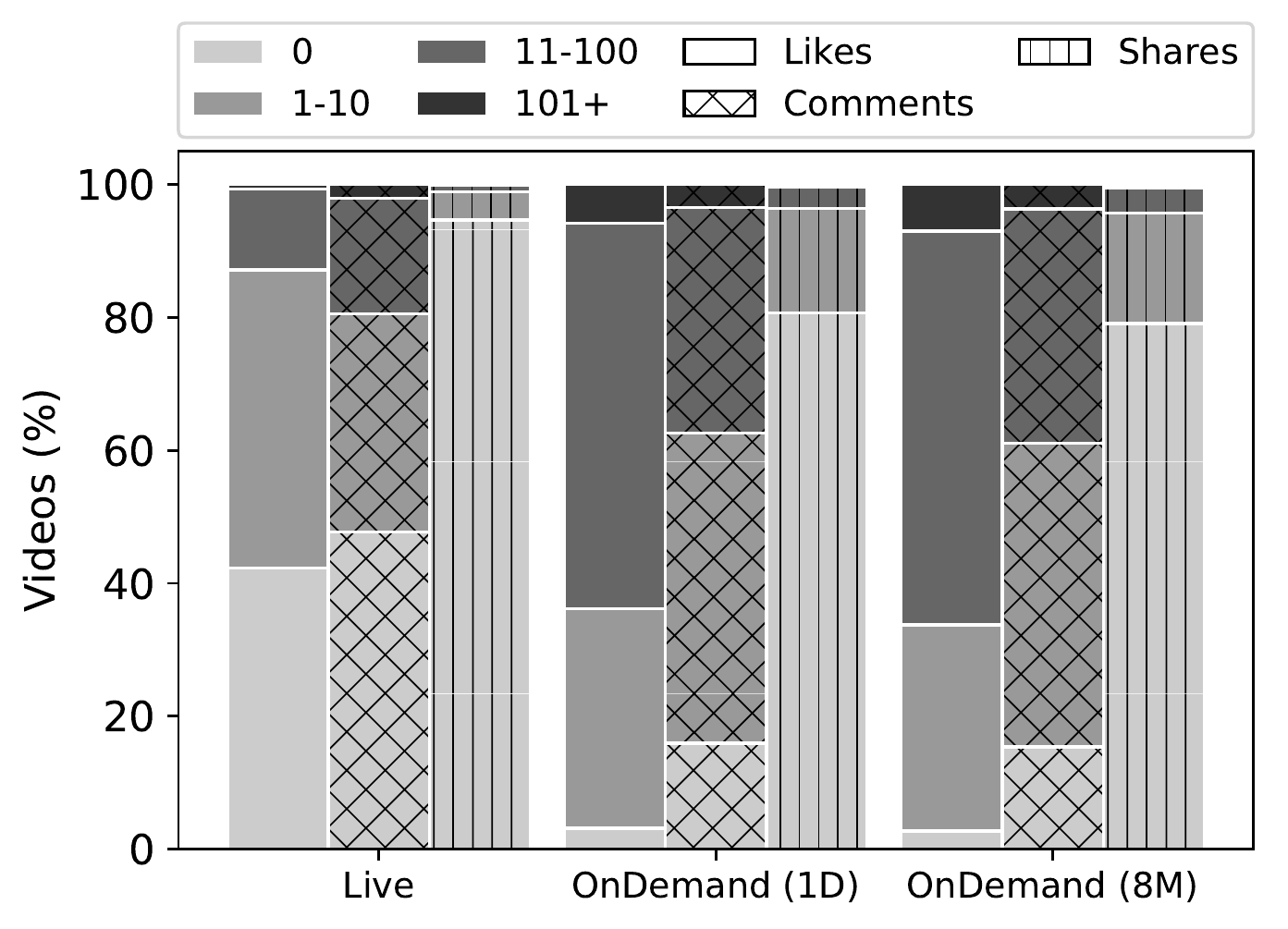}
	\caption{Distribution of engagement metrics (shares, likes, comments) across all broadcast streams.}
	\label{fig:cdf_vids_eng}
	\vspace{-3mm}
\end{figure}

\subsection{Social engagement}
A key feature of Facebook Live is the ability for viewers to interact with broadcasters. This comes in the form of likes, shares and comments. In a sense, this offers an alternative measure of popularity when compared to the earlier discussion of viewer counts. 

Figure~\ref{fig:cdf_vids_eng} presents the distribution of these engagement metrics for all broadcasts. The figure presents these metrics across three time windows: \one~During the live broadcast; \two~a day later for videos that have been archived on-demand; and \three~8 months later. The pattern indicates the type of engagement, whilst the grayscale colour separates broadcasts into four popularity groups -- those that get 0 engagement (likes/comments/shares), 1--10 engagements, 11--100 engagements or more than 100 engagements. 

Most of the engagement happens \emph{after} the live broadcast. It is therefore questionable to what extent this is a \emph{live} platform. On average during the live broadcast, videos receive  6.7 likes, 8.4 comments and 0.54 shares. One day after broadcast, the engagement counts jump to 29.84 likes, 16.33 comments, and 1.33 shares. These numbers do not change much in the next 8 months; thus the majority of interaction happens in the 1 day after broadcast. Notice that about 40\% (47\%) of videos collect no likes (comments) during the live broadcast, but this number drops to 3\% (16\%) after a day. From an infrastructure design perspective, this interaction pattern makes it easier to support social engagement, since on-demand is easier to manager than live streaming, which requires synchronisation of the display of likes and comments across all viewers at the point in the stream where the interaction happened.

We note that during the live broadcasts, comments tend to outnumber likes, but the engagement after the  broadcast tends to be through likes. Comments represent a much stronger form of engagement; thus we conjecture that users who watch and interact with the live stream may be close friends of the user, whereas the later engagement could be from other friends or the general public. \looseness=-1

%% file: sections/conclusions.tex
\section{Conclusion, Implications and Future Work} \label{sec:conclusions}

In this paper we explored the usage of Facebook Live in terms of three major factors: video characteristics; geography of broadcasters and viewers; and the social engagement. We began by characterising broadcast behaviour (\S\ref{sec:char}). Being a popular platform, Facebook Live encounters peaks of 63 million online viewers. Despite this, we find that popularity is typically reserved for Page accounts and a small set of celebrity users. Over half of the social users on Facebook Live generate unwatched content at least once, with 41.5\% of broadcasts never being viewed. This casts doubt over both the terms \emph{social} and \emph{broadcast} in social live broadcast. In response, we proposed a simple mechanism to alleviate network and battery consumption. In cases where live streams receive no viewers, we proposed locally storing the content on the broadcasters' mobile devices until viewers arrive (rather than uploading video chunks that are never watched). Although this may introduce a slightly greater startup delay from viewers, such a model would reduce the number of bytes transferred by 21.9\%, and mean that 23.18\% of videos never need to leave the mobile device (only those that are viewed or archived by the broadcasters, allowing on-demand access later on, must be uploaded).

We then proceeded to explore the geographical (\S\ref{sec:geo}) properties of the system. Due to the social nature of Facebook Live, streams exhibit high degrees of locality. 8\% of viewers are within 25 KM of the broadcaster. We argue that this makes the service well suited to many recently proposed mobile edge computing architectures (\eg~\cite{yuyi2017arxiv,mach2017mobile,raman2017wistitch,adisorn2017picasso}). Unsurprisingly, these observations also result in clear trends in consumption with most users accessing domestic content. Exceptions to this tend to align with high levels of cultural and language similarity. Finally, we explored viewer engagement (\S\ref{sec:social}) to find that even popular streams (\ie $>$10 viewers) have periods of being unwatched. We also found that most social engagement actually occurs \emph{after} the live stream for the 46\% of streams that were archived. This, again, casts doubt over the term \emph{live} in social live broadcast, with Facebook Live exhibiting on-demand-style behaviour for many consumers. 

This paper constitutes a key contribution within the broader field of mobile social video consumption. However, there remains a series of interesting points for exploration. We have noted several systems implications from our findings. Implementing and evaluating these concepts would be a fruitful line of exploration, particularly as many are simple and could be easily integrated into social broadcast apps. We should note, however, that we anticipate significant growth in the use of Facebook Live in the future. Hence, it is important to monitor how behaviour evolves to understand the long-term applicability of design choices. To date, we have also not been able to dive into how the social graph impacts video popularity, nor how such data could be used to predict and inform delivery strategies. Understanding how social topology impacts content consumption would be a clear line of future work.  We also found that a number of spatial properties were correlated; we wish to expand this type of correlation analysis to other domains, \eg understanding if broadcasts correlate with socioeconomic metrics or things like tourism and immigration levels from other countries. Predictive models for capacity planning and/or caching could make great use of such insight.

\begin{acks}
	The work is partially supported by the \grantsponsor{}{EU-India REACH Project}{http://www.eu-india.net/} under Grant ~\grantnum{ICI+/2014/342-986}{ICI+/2014/342-986}, by \grantsponsor{}{UK Engineering and Physical Sciences Research Council (EPSRC)}{} via the Internet of Silicon Retinas Project, Grant No. ~\grantnum{EP/P022723/1}{EP/P022723/1}, an EPSRC Impact Acceleration award, and by the \grantsponsor{}{H2020 MONROE Project}{https://www.monroe-project.eu/} Open Call 2 Grant CaMCoW, under Grant \grantnum{}{No. 644399}.
	
\end{acks}